\documentclass[10pt,letterpaper]{article}
\usepackage{geometry}
\usepackage{graphicx}
\usepackage{setspace}
\usepackage[scaled=0.9]{helvet}
\begin{document}

\title{High-Yield of Memory Elements from Carbon Nanotube Field-Effect Transistors with Atomic Layer Deposited Gate Dielectric}

\author{Marcus~Rinki{\"o}$^{1}$, Andreas~Johansson$^{1,\ast}$, Marina~Y.~Zavodchikova$^{1,2}$, J.~Jussi~Toppari$^{1}$, \\ Albert G. Nasibulin$^{2}$, 
Esko I. Kauppinen$^{2}$, and P{\"a}ivi~T{\"o}rm{\"a}$^{1,2}$ \\
{\footnotesize $^{1}$Department of Physics, Nanoscience Center, P.O. Box 35, 40014 University of Jyv\"{a}skyl\"{a}, FINLAND} \\ 
{\footnotesize $^{2}$Center for New Materials and Laboratory of  Physics, Helsinki University of Technology, P.O. Box 1602, 02044 Espoo, FINLAND} \\
{\footnotesize $^{\ast}$E-mail: andreas.johansson@phys.jyu.fi}}

\twocolumn[
\maketitle

\begin{@twocolumnfalse}

{\bf Carbon nanotube field-effect transistors (CNT FETs) have been proposed as possible building blocks for future nano-electronics. But a challenge 
with CNT FETs is that they appear to randomly display varying amounts of hysteresis in their transfer characteristics. The hysteresis is often 
attributed to charge trapping in the dielectric layer between the nanotube and the gate. This study includes 94 CNT FET samples, providing an 
unprecedented basis for statistics on the hysteresis seen in five different CNT-gate configurations. We find that the memory effect can be 
controlled by carefully designing the gate dielectric in nm-thin layers. By using atomic layer depositions (ALD) of HfO$_{2}$ and TiO$_{2}$ in a 
triple-layer configuration, we achieve the first CNT FETs with consistent and narrowly distributed memory effects in their transfer 
characteristics.}

\vspace*{10pt} 
\end{@twocolumnfalse}
]

The quasi-one-dimensional, nearly perfect, crystalline structures of CNTs make them promising candidates \cite{Dekker1999, Dai2002, McEuen2003, 
Lu2007} to extend the down-scaling of electronic components beyond the limitations of present Si-based technology. Featuring either semiconducting 
or metallic transport properties, they can in principle replace both active components as well as their interconnects. It was suggested that 
CNT-based devices could be mounted with an integration level of up to 10$^{12}$ cm$^{-2}$\cite{Rueckes2000}, which is about 4 orders of magnitude 
higher density than in current technology.    
Field-effect transistors have been made out of CNTs with impressive device parameters, e.g. subthreshold slopes close to 60 mV/decade 
\cite{Weitz2007}, 
and carrier mobilities of up to 9000 cm$^{2}$/Vs \cite{Fuhrer2002}. 
But there are many challenges with incorporating CNTs into logic devices, such as being able to separate the semiconducting CNTs from the metallic 
ones, or to control their placement with nanometer accuracy. 
Another challenge with CNT FETs is that often they display some degree of hysteresis in their transfer characteristics. For a CNT FET this is an 
unwanted feature, rendering it unpredictable in its output, and it has motivated several studies to find ways to prevent or remove these tendencies 
\cite{Weitz2007, Kim2003, Ozel2005, Wang2005, McGill2006}. On the other hand, the presence of hysteresis opens up the possibility to utilize the 
device as a memory element instead. This has been pointed out by several studies \cite{Fuhrer2002, Radosavljevic2002, Cui2002, Choi2003, 
Bradley2003, Wang2005b}, demonstrating CNT FETs with ON and OFF states which are well separated and addressable with positive or negative gate 
voltage pulses. However, the challenge is to be able to control the presence of hysteresis, which so far has been reported as a more or less random 
property among the studied CNT FETs \cite{Cui2002, Bradley2003, Vijayaraghavan2006}. In this letter, we use Hf-based ALD-grown gate dielectric to 
control hysteresis and achieve a 100 \% yield in memory effect, as well as study the origin of the hysteresis in our devices. 
This is to our knowledge the first report on CNT FETs featuring {\it consistent} memory effects.

Several different models have been suggested to explain the hysteresis in the transfer characteristics of a CNT FET. 
First it was pointed out that, especially for CNT FETs with a gate insulator of SiO$_{2}$, it may have the same origin as the hysteresis sometimes 
seen in conventional Si-MOSFETs \cite{Fuhrer2002, Radosavljevic2002, Bradley2003}. There it was shown that mobile ions or charges within the 
SiO$_{2}$ layer could relocate in response to the applied gate voltage, and as a result modify the local electric field sensed by the charge 
carriers in the conduction channel. But this is not the only proposed mechanism. 
It has also been suggested \cite{Fuhrer2002, Robert-Peillard2005, Kar2006} that the charge traps may not be mobile, but located stationary in the 
near vicinity of the CNT. An applied gate voltage could then assist in filling or emptying the charge traps with charge carriers moving in the CNT, 
which in turn screens the applied electric field and causes hysteresis to appear in the gate voltage response. 
Yet another possibility is that surface chemistry plays an important role. E.g. water molecules adhered to the surface have been shown 
\cite{Kim2003,Wang2005} to give a large contribution to the hysteresis of some CNT FETs. 
A fourth model suggests that defects in the nanotube itself could provide charging centers, which can be filled or emptied in response to the gate 
modulation. A charging center in this case may be a carbon atom substituted with a different atom or molecule, which can donate or accept electrons 
from the conduction channel.

\begin{figure}[h]
\includegraphics[width=7.5cm]{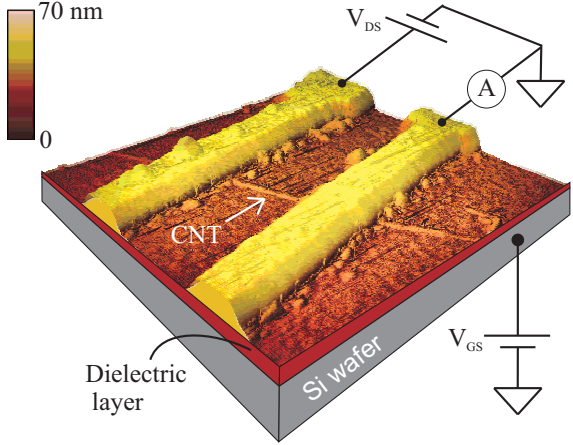}
\caption{{\bf AFM image of a typical device, covering an area of 1 $\mu$m$^{2}$.} The SWCNT is resting on the dielectric layer, and have source and 
drain electrodes of Pd deposited on top with a spacing of 260 nm. The measurement setup is schematically drawn with voltage applied to the Si wafer, 
acting as a backgate, and voltage applied between the drain and source electrodes while measuring the the current response through the CNT.} 
\label{MOME2_overview}
\end{figure}

This study includes in total 94 SWCNT FETs with varying gate insulator, which by far exceeds the number of samples included in earlier studies of 
hysteresis in CNT FETs \cite{Fuhrer2002, Kim2003, Radosavljevic2002, Bradley2003, Wang2005, Choi2003, Robert-Peillard2005, Kar2006}. An atomic force 
microscope (AFM) image of a typical sample is shown in Fig. \ref{MOME2_overview}. The samples were built in a bottom-up approach, described in 
detail in the Methods section and ref. \cite{Zavodchikova2007}.
A schematic of the measurement setup is drawn in Fig. 1. Measurements were carried out at ambient conditions, and are described in the Methods 
section.     
17 of the 94 samples had the backgate covered with ALD of Al$_{2}$O$_{3}$ (nominal thickness: 20 nm), 14 with an ALD of HfO$_{2}$ (20 nm), 11 with 
an ALD of HfO$_{2}$-TiO$_{2}$-HfO$_{2}$ (40-0.5-3 nm), 27 with an ALD of HfO$_{2}$-TiO$_{2}$-HfO$_{2}$ (40-0.5-1 nm), and the remaining 25 with 
thermally grown SiO$_{2}$ (300 nm). 

\begin{figure}[t,h]
\includegraphics[width=7.5cm]{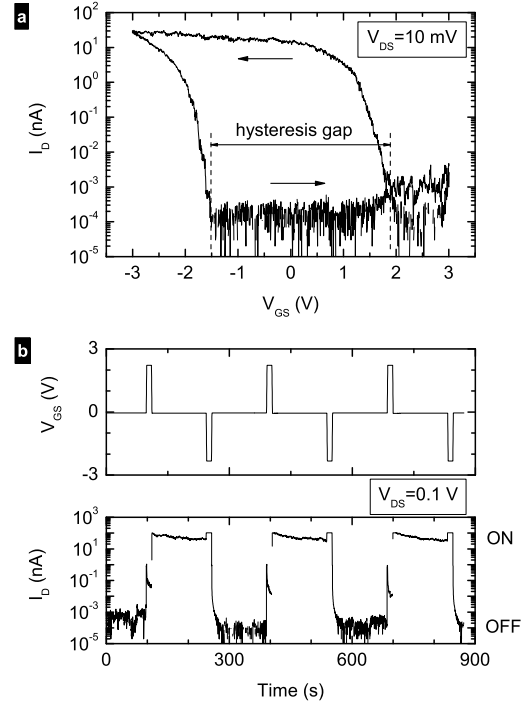}
\caption{{\bf Memory effect in CNT FET.} {\bf a,} Drain current versus backgate voltage for a typical SWCNT FET with a gate dielectric of HfO$_{2}$. 
The arrows mark the scan direction within the loop. The hysteresis gap is given by the difference in threshold voltage between the reverse and the 
forward scan direction. {\bf b,} Demonstration of its memory function. The upper pane displays the voltage applied to the back gate as a function of 
time. The lower pane shows how  the current response through the CNT switches to an ON state or an OFF state in response to a positive or negative 
voltage pulse on the backgate, respectively.
}
\label{MOME2_hysteresis}
\end{figure}

The focus of this study is on the appearance of hysteresis in the transfer characteristics of our CNT FETs. A typical example is shown in Fig. 
\ref{MOME2_hysteresis}a for a SWCNT resting on a backgate dielectric of 20 nm thick HfO$_{2}$ with 10 mV applied between the drain and source 
electrodes. Most SWCNT FETs displayed typical unipolar p-type behavior \cite{Tans1998a, Martel1998}, with strongly suppressed conductance at 
positive gate voltages and a transition into a highly conducting state at negative gate voltages. A few devices showed ambipolar dependence with a 
somewhat increased conductance at high positive backgate voltages, which has been attributed to semiconducting nanotubes with a small bandgap 
\cite{Martel2001, Avouris2002}. Upon scanning the backgate voltage back and forth, the threshold voltage attains in some, but not all, of the CNT 
FETs a higher value for the reverse sweep than for the forward sweep, resulting in a highly reproducible hysteresis loop with different conductance 
values at zero backgate voltage depending on the sweep direction. In Fig. \ref{MOME2_hysteresis}b it is demonstrated how the current response of 
such a CNT FET can be switched between the two states by sending either a positive or negative voltage pulse to the backgate.

All mass-fabricated electronic devices have a natural variation of characteristic parameters, which is acceptable as long as the parameter 
distribution is narrow enough not to interfere with its intended function. The large number of devices in this study allows us to estimate the 
distribution of hysteresis response seen for differing gate dielectrics. We quantify the memory effect in each device in terms of the shift in 
threshold voltage, called the hysteresis gap (see Fig. \ref{MOME2_hysteresis}a). This measure relates directly to the reconfiguration of charges 
trapped in the close vicinity of the SWCNT, and is sensitive to the gate voltage scan rate, the scan range, as well as the hold time at the turning 
points of the scanning interval before starting the next scan. While the scan rate was kept at 10 mV/s and the hold time at the turning points was 
close to 1 second throughout the study, the gate voltage scan range was altered between the samples due to differing gate insulator thicknesses. The 
samples with 300 nm thick SiO$_{2}$ dielectric were measured with a gate voltage scan range of $\pm$ 10 V, while the ALD based samples with gate 
insulator thicknesses of 20--43.5 nm had gate voltage scan ranges of $\pm$ 2--3 V. To allow comparison between these different cases, we calculated 
the relative hysteresis gap, where the hysteresis gap is normalized by the gate scan range. 

\begin{figure}[h]
\includegraphics[width=7.5cm]{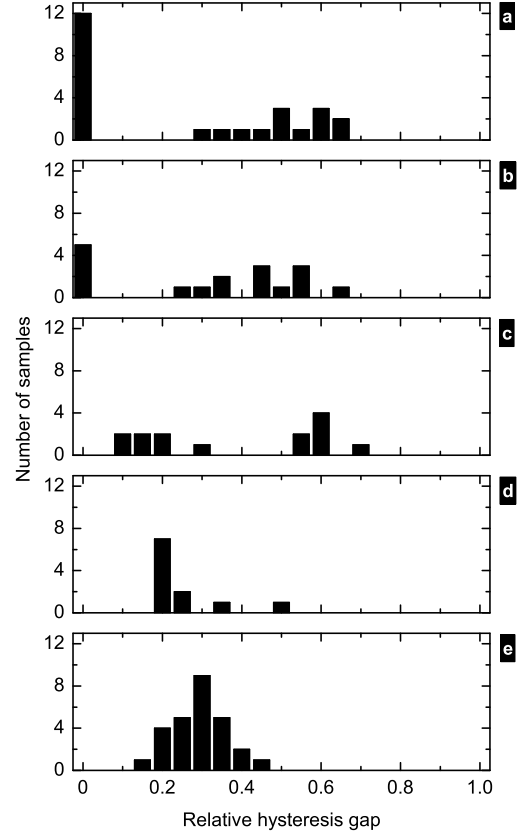}
\caption{ {\bf Statistics of the relative hysteresis gap from 94 devices with varying gate dielectric.} Each column represents a 5 \% interval of 
the full scale. The gate dielectric is in {\bf a,} SiO$_{2}$ (300 nm), {\bf b,} Al$_{2}$O$_{3}$ (20 nm), {\bf c,} HfO$_{2}$ (20 nm), {\bf d,} 
HfO$_{2}$-TiO$_{2}$-HfO$_{2}$ (40-0.5-3 nm), and {\bf e,} HfO$_{2}$-TiO$_{2}$-HfO$_{2}$ (40-0.5-1 nm).}
\label{Rel_Hyst_Gap}
\end{figure}

Our results are plotted in Fig. \ref{Rel_Hyst_Gap}. Each column represents a 5 \% interval of the relative hysteresis gap along the x-axis. We show 
in Fig. \ref{Rel_Hyst_Gap}a that from 25 SiO$_{2}$-based CNT FETs, 12 devices do not exhibit hysteresis at all. The remaining 13 display a relative 
hysteresis gap almost evenly spread within the interval 30--65 \%. The result is in agreement with earlier reports \cite{Cui2002,Bradley2003} on 
SiO$_{2}$-based SWCNT FETs, finding that only a fraction of the produced devices show hysteresis in their transfer characteristics. 
The picture is very similar in Fig. \ref{Rel_Hyst_Gap}b, where the SWCNT FETs have a gate dielectric of 20 nm thick ALD grown Al$_{2}$O$_{3}$. 5 
devices show no hysteresis while 12 devices have their relative hysteresis gap within the interval 25--65 \%.  

For the remaining three panes of Fig. \ref{Rel_Hyst_Gap}, there is a distinct difference in that {\it all} the fabricated devices display a clear 
relative hysteresis gap. With memory devices in mind, this translates into a 100 \% fabrication yield. In Fig. \ref{Rel_Hyst_Gap}c, the gate 
dielectric is 20 nm thick ALD grown HfO$_{2}$. The 14 devices made have their relative hysteresis gaps spread from 10 \% up to 70 \% of the total 
gate scan range. While all CNT FETs in this set show a memory effect, the distribution of the relative hysteresis gaps is very wide, having a 
standard deviation of 22 \%. 
The following two panes of Fig. \ref{Rel_Hyst_Gap} show data from a specially designed three-layered ALD structure. It was made in order to study 
the hypothesis that the lower interface of the top most HfO$_{2}$ layer plays an important role in controlling the amount of charge traps available. 
The first deposited layer is a 40 nm thick buffer layer of HfO$_{2}$ with the purpose of providing a stable dielectric which can withstand the 
physical stress of sample processing. Next a 0.5 nm thick layer of TiO$_{2}$ was deposited with the intent of creating an interface to the top most 
layer. Finally another layer of HfO$_{2}$ was deposited, with a thickness of 3 nm in Fig. \ref{Rel_Hyst_Gap}d and 1 nm in Fig. \ref{Rel_Hyst_Gap}e. 
While the total thicknesses of the ALD dielectrics in Figs. \ref{Rel_Hyst_Gap}d and \ref{Rel_Hyst_Gap}e are about twice the thickness in Fig. 
\ref{Rel_Hyst_Gap}c, the lower interface of the top HfO$_{2}$ layers are moved considerably closer to the CNT. The resulting relative hysteresis gap 
distributions show strongly decreased standard deviations of 0.09 and 0.06 for Fig. \ref{Rel_Hyst_Gap}d and Fig. \ref{Rel_Hyst_Gap}e, respectively. 
More so, the latter pane with a top layer of only 1 nm thickness displays a distribution of relative hysteresis gap values that closely resemble the 
normal distribution. This indicates that the distribution is not likely to change noticeably if we were to add more samples to the study.         

\begin{table}[htb]
\caption{{\bf Statistics taken from data in Fig. \ref{Rel_Hyst_Gap}, with the lettering of gate dielectric in the first column according to that of 
the panes. The following columns display the total number of samples, the mean relative hysteresis gap, and its standard deviation.}}
\begin{center}
\begin{tabular}{cccc}
	\hline 
  	gate & no. of & mean rel. & standard \\
  	dielectric & samples & hyst. gap & deviation \\ \hline  
  	a  & 25 & 0.27 & 0.27 \\
  	b  & 17 & 0.32 & 0.24 \\
  	c  & 14 & 0.38 & 0.22 \\
  	d  & 11 & 0.26 & 0.09 \\
  	e  & 27 & 0.29 & 0.06 \\
\end{tabular}
\end{center}
\label{Statistics}
\end{table}

Our results are summarized in Table \ref{Statistics}, displaying the number of samples, the mean of the relative hysteresis gap, and its standard 
deviation for each type of gate dielectric according to the lettering in Fig. \ref{Rel_Hyst_Gap}. Clearly, the triple layer with thinner upper 
HfO$_{2}$ layer is the better choice as gate insulator when preparing a memory storage device. More important, we show here that memory effects in a 
CNT FET can be controlled in ambient conditions, even without applying any kind of surface passivation layer on to the device. To test the influence 
of surface chemistry (e.g. adhered H$_{2}$O molecules) on the memory effects seen, 12 samples with the thinner triple-layer ALD dielectric were 
measured in a chamber with dry nitrogen gas flow, reducing the relative humidity to less than 1 \%. The relative hysteresis gap was not affected for 
about half of the samples, while the other half saw a slight decrease. The changes in relative hysteresis gap were in all cases less than 10\%. The 
ON and OFF states remained unaffected. It is therefore reasonable to assume that molecules adhered to the surface play a minor role in the memory 
effects seen. 

The fact that changes in memory effects follow changes in gate dielectric, with relatively narrow distributions in two cases, discredit the model 
that defects in the SWCNTs are providing the charge traps responsible for the hysteresis. The reason for the differences seen between single layer 
dielectrics of SiO$_{2}$, Al$_{2}$O$_{3}$ and HfO$_{2}$ is difficult to discern, but may be related to differing charge trap densities, or even 
types of charge defects.
     
Turning our attention to the rapidly narrowing distribution in hysteresis gap when going from single layer HfO$_{2}$ to a triple-layered dielectric 
structure, the most significant difference in the surrounding of the CNT is the closely located interface between the HfO$_{2}$ and the TiO$_{2}$ 
layer. It is common knowledge that the interface between two different materials may carry defects or charge traps. An alternative explanation could 
be that as the top-layer of HfO$_{2}$ becomes thinner, its surface becomes rougher and hosts an increasing number of defects, e.g. oxygen vacancies. 
The defects may act as charge traps and cause the increased hysteresis. But here we would like to point out two opposing observations. First, 
surface defects are likely to interact with or be screened by adhered water molecules, which then would decrease the relative hysteresis gap in 
ambient conditions. We see the opposite trend, a slight decrease of the relative hysteresis gap in vacuum. Second, our AFM measurements show no 
quantitative difference in surface roughness between the samples with different layer thicknesses. Here should be added though that the AFM is 
working close to its limit of resolution, measuring roughesses on the order of 1 nm.
While mobile charge traps also may be present within the dielectric, and could possibly be a major contributor to memory effects seen in the single 
layer dielectrics, it is reasonable to assume on the basis of these data that moving the lower interface of the upper HfO$_{2}$ layer closer to the 
CNT provides a layer of stationary charge traps at a well calibrated distance from the CNT. It is supported by the notion that a layer of stationary 
charge traps in the vicinity of the CNT will also screen the action from mobile charge carriers, thus creating a well defined device geometry with a 
narrow hysteresis gap distribution.
We therefore conclude that the most probable dominating charge storage mechanism in the triple layer structure is due to stationary charge traps at 
the lower interface of the uppermost HfO$_{2}$ layer, which are filled and emptied by charge carriers in response to application of a negative or 
positive gate voltage.      

In summary, we have investigated 94 SWCNT FETs with different gate insulators, giving us unprecedented statistics on the presence of hysteresis in 
their transfer characteristics. We find that by using a tree-layered gate dielectric consisting of HfO$_{2}$-TiO$_{2}$-HfO$_{2}$, all SWCNT FETs 
display hysteresis with a narrow distribution of their relative hysteresis gaps, which narrows down further when the upper most layer is changed 
from 3 nm to 1 nm. The study shows that SWCNT FETs can be fabricated and operated in ambient conditions without any surface passivation, which 
points towards the dominant charge trapping mechanism being insensitive to H$_{2}$O molecules on the surface. Such layered gate dielectric is of 
particular interest for memory applications, providing to our knowledge the first proven route to 100 \% yield in single CNT-based memory elements. 
However, several challenges remain to be solved in order to make highly integrated CNT memory cells, such as positioning the CNTs with nm precision 
and providing each of them with a local, nanotube specific gate.  
\vspace*{10pt}

{\bf Methods}

The samples were prepared by starting from a highly boron-doped Si wafer which acts as a backgate. The backgate was covered with an ALD layer of 
either Al$_{2}$O$_{3}$ (nominal thickness: 20 nm), HfO$_{2}$ (20 nm), or HfO$_{2}$-TiO$_{2}$-HfO$_{2}$ (40-0.5-3 or 40-0.5-1 nm). All ALD 
depositions were done at Planar Systems Inc. (Espoo, Finland), using a Planar P400A ALD deposition tool. The idea motivating the triple-layer 
structure is to create, in a controlled way, charge traps in the close vicinity of the CNT FET \cite{Choi2003}. The interface between two different 
ALD layers may serve that purpose. We also prepared reference samples with gate insulator of the more commonly used thermally grown SiO$_{2}$ (300 
nm). 
On top of the gate insulator a matrix of alignment markers was deposited, using e-beam lithography and metallization of Pd with an adhesion layer of 
Ti. CNTs were then deposited in two different ways, depending on their origin. 
Our primary source was commercial SWCNTs produced by NanoCyl S.A. (Sambreville, BELGIUM), bought in the shape of black powder, which was suspended 
in a solvent by ultra-sonication. A few droplets of the nanotube suspension were then deposited onto the sample. We also prepared, as a reference, 
11 samples with SWCNTs from the Hot Wire Generator reactor on to substrates with the HfO$_{2}$-TiO$_{2}$-HfO$_{2}$ (40-0.5-1 nm) triple-layer. The 
Hot Wire Generator reactor is based on the aerosol (floating catalyst) synthesis of CNTs. Iron particles and carbon monoxide were utilized as the 
catalyst and the carbon source, respectively \cite{Nasibulin2005}. Individual CNTs were filtered out from the bundled tubes in the gas phase as 
described in \cite{Gonzalez2006} and deposited onto the sample surface at room temperature using thermophoretic precipitator \cite{Gonzalez2005}. 
In both methods CNTs were left at random places on the surface. Their locations were then mapped in relation to the matrix of alignment markers, 
using an AFM. Finally, electrodes of Pd were deposited onto the ends of the CNTs, with the help of e-beam lithography and subsequent metallization. 

All measurements in this study were carried out at room temperature in an electrically shielded room, either under ambient conditions or in a 
chamber under dry nitrogen gas flow, which reduced the relative humidity to below 1\% (below the resolution of our humidity sensor). Two-terminal 
measurements were performed, with the substrate acting as a backgate. DC measurements were used with the applied voltage given by a home-built 
voltage distribution box, powered by batteries and computer controlled via a data acquisition card, while measuring the current response through the 
nanotube. I-V characteristics and conductance response to an applied backgate voltage was collected from all samples. 
The samples with linear I-V characteristics and its conductance not sensitive to an applied backgate voltage were considered to have metallic CNTs. 
The study includes in total 94 semiconducting SWCNTs, featuring a clear backgate dependence. These comprise about 82 \% of all the samples made, 
which is somewhat higher than the expected 67 \% for randomly picked CNT chiralities. Surprisingly, eight of the eleven devices made with SWCNTs 
from the Hot Wire Generator reactor show clear metallic behavior. Of the three remaining semiconducting SWCNTs, one had a malfunctioning back-gate, 
leaving only two CNT FETs of this kind added to the study. As shown in Fig. 3, these two CNT FETs do not deviate notably in performance compared to 
the other 25 devices with the same gate dielectric.  

An often problematic part in CNT sample processing is to achieve low contact resistance between electrode and nanotube \cite{Chen2005}. As an upper 
limit measure of the contact resistances in our devices, we sampled the total two-terminal resistances of the metallic CNTs. These were found to be 
in the range of 14-160 k$\Omega$, which is close to the theoretical minimum resistance for SWCNTs of $1/2G_{0} = h/4e^2 \approx 6.45 k \Omega$. Here 
G$_{0}$ is the quantum unit of conductance, e is the charge of the electron and h is Planck's constant.

The memory devices included in this study have so far been subjected to slow switching frequencies of up to 10 Hz, using typically gate voltages of 
$\pm$ 3 V for ALD gated samples and $\pm$ 10 V for CNT FETs with gate dielectric of SiO$_{2}$. Some of the devices show charge stability with no 
change in ON or OFF state for several days, while others have a retention time of down to a few hours. Durability has been tested for some of the 
CNT memories, with no significant change in ON/OFF states after switching 10$^{4}$ times or more. We are currently studying these aspects, but would 
like to add they should not be compared to single device characteristics of state-of-the-art commercial memories before the gate configuration is 
changed from a backgate to a local gate for each CNT, which may significantly improve device characteristics. That work is also in progress. 

{\bf Acknowledgement}
We acknowledge valuable discussions with Prof. H. H{\"a}kkinen, and our research project collaborators: Prof. M. Ahlskog's and Prof. K. Rissanen's 
groups at University of Jyv{\"a}skyl{\"a}, as well as the companies Vaisala Oyj and Nokia Oyj. We thank Planar Inc., Olli Jylh{\"a} for ALD 
processing and suggesting the materials used. This work was supported by the Finnish Funding Agency for Technology and Innovation (TEKES), project 
number 40309/05, and the Academy of Finland. M. R. would like to acknowledge support from the Magnus Ehrnrooth Foundation and the Tekniikan 
Edist{\"a}miss{\"a}{\"a}ti{\"o} (TES).


\begin{thebibliography}{10}

\bibitem{Dekker1999}
Dekker,~C. Carbon nanotubes as molecular quantum wires. \textit{Phys. Today} \textbf{52,} 22--28 (1999).

\bibitem{Dai2002}
Dai, H. Carbon nanotubes: opportunities and challenges. \textit{Surf. Sci.} \textbf{500,} 218--241 (2002).

\bibitem{McEuen2003}
McEuen, P.~L., Fuhrer, M.~S. \& Park, H.~K. Single-walled carbon nanotube electronics. \textit{IEEE Trans. Nanotechnology} \textbf{1,} 78--85 
(2003).

\bibitem{Lu2007}
Lu, W. \& Lieber, C.~M. Nanoelectronics from the bottom up. \textit{Nature Mater.} \textbf{6,} 841--850 (2007).

\bibitem{Rueckes2000}
Rueckes, T. \emph{et~al.} Carbon nanotube-based nonvolatile random access memory for molecular computing. \textit{Science} \textbf{289,} 94--97 
(2000).

\bibitem{Weitz2007}
Weitz, R.~T. \emph{et~al.} High-performance carbon nanotube field effect transistors with a thin gate dielectric based on a self-assembled 
monolayer. \textit{Nano Lett.} \textbf{7,} 22--27 (2007).

\bibitem{Fuhrer2002}
Fuhrer, M.~S., Kim, B.~M., Durkop, T. \& Britlinger, T. High-mobility nanotube transistor memory. \textit{Nano Lett.} \textbf{2,} 755--759 (2002).

\bibitem{Kim2003}
Kim, W. \emph{et~al.} Hysteresis caused by water molecules in carbon nanotube field-effect transistors. \textit{Nano Lett.} \textbf{3,} 193--198 
(2003).

\bibitem{Ozel2005}
Ozel, T., Gaur, A., Rogers, J.~A. \& Shim, M. Polymer electrolyte gating of carbon nanotube network transistors. \textit{Nano Lett.} \textbf{5,} 
905--911 (2005).

\bibitem{Wang2005}
Wang, S., Sellin, P., Zhang, Q. \& Yang, D. Nonvolatile memory from single-walled carbon nanotube-based field effect transistors. \textit{Current 
Nanoscience} \textbf{1,} 43--46 (2005).

\bibitem{McGill2006}
McGill, S.~A., Rao, S.~G., Manandhar, P., Xiong, P. \& Hong, S. High-performance, hysteresis-free carbon nanotube field-effect transistors via 
directed assembly. \textit{Appl. Phys. Lett.} \textbf{89}, 163123--163125 (2006).

\bibitem{Radosavljevic2002}
Radosavljevic, M., Freitag, M., Thadani, K.~V. \& Johnson, A.~T. Nonvolatile molecular memory elements based on ambipolar nanotube field effect 
transistors. \textit{Nano Lett.} \textbf{2,} 761--764 (2002).

\bibitem{Cui2002}
Cui, J.~B., Sordan, R., Burghard, M. \& Kern, K. Carbon nanotube memory devices of high charge storage stability. \textit{Appl. Phys. Lett.} 
\textbf{81,} 3260--3262 (2002).

\bibitem{Choi2003}
Choi, W.~B. \emph{et~al.} Carbon-nanotube-based nonvolatile memory with oxide-nitride-oxide film and nanoscale channel. \textit{Appl. Phys. Lett.} 
\textbf{82,} 275--277 (2003).

\bibitem{Bradley2003}
Bradley, K., Cumings, J., Star, A., Gabriel, J.-C.~P. \& Gruner, G. Influence of mobile ions on nanotube based fet devices. \textit{Nano Lett.} 
\textbf{3,} 639--641 (2003).

\bibitem{Wang2005b}
Wang, S. \& Sellin, P. Pronounced hysteresis and high charge storage stability of single-walled carbon nanotube-based field-effect transistors. 
\textit{Appl. Phys. Lett.} \textbf{87,} 133117--133119 (2005).

\bibitem{Vijayaraghavan2006}
Vijayaraghavan, A. \emph{et~al.} Charge-injection-induced dynamic screening and origin of hysteresis in field-modulated transport in single-wall 
carbon nanotubes. \textit{Appl. Phys. Lett.} \textbf{89}, 162108--162110 (2006).

\bibitem{Robert-Peillard2005}
Robert-Peillard, A. \& Rotkin, S.~V. Modeling hysteresis phenomena in nanotube field-effect transistors. \textit{IEEE Trans. Nanotechnol.} 
\textbf{4,} 284--288 (2005).

\bibitem{Kar2006}
Kar, S. \emph{et~al.} Quantitative analysis of hysteresis in carbon nanotube field-effect devices. \textit{Appl. Phys. Lett.} \textbf{89,} 
132118--132120 (2006).

\bibitem{Zavodchikova2007}
Zavodchikova, M.~Y. \emph{et~al.} Fabrication of carbon nanotube-based field-effect transistors for studies of their memory effects. \textit{phys. 
stat. sol. (b)} \textbf{244,} 4188--4192 (2007).

\bibitem{Nasibulin2005}
Nasibulin, A.~G., Moisala, A., Brown, D.~P., Jiang, H. \& Kauppinen, E.~I. A novel aerosol method for single walled carbon nanotube synthesis. 
\textit{Chem. Phys. Lett.} \textbf{402,} 227--232 (2005).

\bibitem{Gonzalez2006}
Gonzalez, D. \emph{et~al.} Spontaneous charging of single-walled carbon nanotubes: A novel strategy for the selective substrate deposition of 
individual tubes at ambient temperature.
\textit{Chemistry of Materials} \textbf{18,} 5052--5057 (2006).

\bibitem{Gonzalez2005}
Gonzalez, D. \emph{et~al.} A new thermophoretic precipitator for collection of nanometer-sized aerosol particles. \textit{Aerosol Sci. Technol.} 
\textbf{39,} 1064--1071 (2005).

\bibitem{Chen2005}
Chen, Z., Appenzeller, J., Knoch, J., Lin, Y. \& Avouris, P. The role of metal-nanotube contact in the performance of carbon nanotube field-effect 
transistors. \textit{Nano Lett.} \textbf{5,} 1497--1502 (2005).

\bibitem{Tans1998a}
Tans, S., Dekker, C. \& Verschueren, A. Room-temperature transistor based on a single carbon nanotube. \textit{Nature} \textbf{393,} 49--52 (1998).

\bibitem{Martel1998}
Martel, R., Schmidt, T., Shea, H., Hertel, T. \& Avouris, P. Single- and multi-wall carbon nanotube field-effect transistors. \textit{Appl. Phys. 
Lett.} \textbf{73,} 2447--2449 (1998).

\bibitem{Martel2001}
Martel, R. \emph{et~al.} Ambipolar electrical transport in semiconducting single-wall carbon nanotubes. \textit{Phys. Rev. Lett.} \textbf{87,} 
256805--256808 (2001).

\bibitem{Avouris2002}
Avouris, P. Molecular electronics with carbon nanotubes. \textit{Acc. Chem. Res.} \textbf{35,} 1026--1034 (2002).

\end{thebibliography}
\end{document}